\documentclass[final,5p,times,twocolumn]{elsarticle}

\usepackage{graphicx}
\usepackage{amssymb}
\usepackage{amsmath,bm}

\usepackage{lineno}
\usepackage{color}
\usepackage{url}
\usepackage{float}
\usepackage{listings}
\usepackage[table,xcdraw]{xcolor}
\usepackage{booktabs}
\usepackage{hyperref}

\journal{Journal Name}

\begin{document}

\begin{frontmatter}


\title{Quantum and Hybrid Machine-Learning Models for Materials-Science Tasks}




\author{Leyang Wang}
\address{$^1$New York University, New York, NY 10003, United States}
\author{Yilun Gong}
\address{$^{2}$Max-Planck-Institut f\"ur Nachhaltige Materialien GmbH, D\"usseldorf, 40237, Germany} 
\address{$^{3}$Department of Materials, University of Oxford, Parks Road, Oxford, OX1 3PH United Kingdom}
\author{Zongrui Pei}
\address{$^1$New York University, New York, NY 10003, United States}
\ead{zp2137@nyu.edu;peizongrui@gmail.com} 

\begin{abstract}
Quantum computing has become increasingly practical in solving real-world problems due to advances in hardware and algorithms. In this paper, we aim to design and estimate quantum machine learning and hybrid quantum-classical models in a few practical materials science tasks, i.e., predicting stacking fault energies and solutes that can ductilize magnesium. To this end, we adopt two different representative quantum algorithms, i.e., quantum support vector machines (QSVM) and quantum neural networks (QNN), and adjust them to our application scenarios. We systematically test the performance with respect to the hyperparameters of selected ansatzes. We identify a few combinations of hyperparameters that yield validation scores of approximately 90\% for QSVM and hybrid QNN in both tasks. Eventually, we construct quantum models with optimized parameters for regression and classification that predict targeted solutes based on the elemental volumes, electronegativities, and bulk moduli of chemical elements.



\end{abstract}


\end{frontmatter}



\section{Introduction}
Quantum information science is one of the most active research domains at the intersection of physics, computer science, and applied mathematics. A few critical research directions in this domain include (i) identifying new realizations of qubits (topological qubits, superconducting qubits, neutral atoms, trapped ions, magnetic properties of defects like nitrogen vacancies in diamond), (ii) development of error-correction algorithms and methods that enhance the fidelity of qubits, (iii) the accurate control of qubit states (realization of various quantum gates and measurements), (iv) formulation of the fundamental theorems (i.e., no-go theorems, quasi-probability), etc.

The mechanism of classical computers is built upon Boolean logic, and its fundamental building block is known as a flip-flop \cite{flip-flop}. 
The flip-flop has two stable states, which can be read as high or low. Such a form of retaining information has been integrated into almost every electronic device. Although the functionality and architecture have undergone significant changes since its invention, the basic computing units still store binary information.
In the 20th century, many scientists had already developed the concept of quantum computing. One well-known source introduces the idea of simulating physics with a computer that exactly captures the physical property of the real world, as opposed to the classical approximations on classical computers. Richard Feynman explained that simulating quantum physics with classical computers was challenging due to the exponential growth in computing resources required \cite{feynman2018simulating,feynman1986quantum}. A computer that is capable of storing probabilistic states will overcome the resource constraint. These ideas that Feynman proposes closely align with the current quantum computers.

One of the major hardware components of quantum computers is quantum bits or qubits, which can be realized by various physical entities \cite{confalone2025cuprate}, such as photons \cite{aghaee2025scaling}, trapped ions \cite{liu2025certified}, superconductors \cite{wilen2021correlated}, bosons \cite{putterman2025hardware}, etc. Each type of qubit has its advantages and disadvantages in terms of scalability, operating temperature, decoherence time, and other factors.
There has been a surge of updates on the development of quantum hardware from research institutions as well as technology companies that have traditionally built their products using classical computing. This marks a growing interest and greater confidence that quantum computing will serve essential roles in the future.
For example, a Google team constructed their quantum computer, Willow, with 105 superconducting qubits \cite{acharya2024quantum}. King {\it et al.} implemented coherent quantum annealing in a programmable 2,000 qubit Ising chain \cite{king2022coherent} and performed quantum critical dynamics in a 5,000-qubit programmable spin glass \cite{king2023quantum}. Recently, the team demonstrated supremacy in a practical physical problem, specifically the dynamics of two-, three-, and infinite-dimensional spin glasses, unlike previous studies that focused on idealized systems \cite{king2025beyond}. In parallel, a Chinese team has transmitted quantum-encrypted images of a record 12,900 kilometers \cite{li2025microsatellite}. 



Machine learning and quantum computers are two technologies that will likely fundamentally change our future. Combining the two, quantum machine learning (QML) has emerged as a novel method with applications in various research domains \cite{peral2024systematic,pei2024designing,pei2024computer,del2025comparative,liu2023machine} (see the literature review of QML from 2017 to 2023 \cite{peral2024systematic}). For example, QML has the potential to revolutionize language models \cite{pei2025language,pei2024towards}, and there are already a few initial works to explore this potential \cite{buonaiuto2024quantum,del2025comparative}. QML algorithms can be grouped into hybrid quantum/classical algorithms and purely quantum algorithms. Hybrid algorithms, such as quantum support vector machines (QSVM), utilize a quantum circuit as a kernel to replace classical kernels. This group of algorithms leverages both classical and quantum hardware, thereby benefiting from the strengths of both. In contrast, purely quantum algorithms, such as quantum neural networks (QNN), rely solely on quantum hardware. Recent theoretical works show that QNNs exhibit similar behavior to their classical counterparts, which may explain their performance; specifically, they form Gaussian processes \cite{garcia2025quantum}. Both hybrid and quantum models can be optimized using classical methods, which is preferred since they are more stable and consistent toward the optimal solutions. Quantum optimization algorithms, such as quantum gradient descent, exhibit probabilistic behavior. Several studies have already implemented QML models to solve real-world problems, such as QSVM models for alloy design \cite{pei2024designing} and QNN models for supply chain logistics \cite{correll2023quantum}. 

In this study, we choose the hybrid QSVM and estimator quantum neural network as examples to build a quantum model for a practical materials-science problem.  
As an example, we show how to use QML to solve a typical and important materials-science problem, i.e., how to predict the stacking fault energies in hexagonal close-packed magnesium alloys \cite{pei2013ab,sandlobes2014ductility,pei2015rapid,pei2019machine}. More specifically, we focus on the intrinsic I$_1$ stacking faults that form as a result of the dissociation of non-basal dislocations. The performance of these algorithms and models relies on highly optimized quantum circuits. We will focus on a parametric comparison and optimization of these models. We will optimize the quantum models in both classification and regression tasks.










\section{Computational details}
\subsection{Methods}
\begin{figure}
    \centering
    \includegraphics[width=1.0\linewidth]{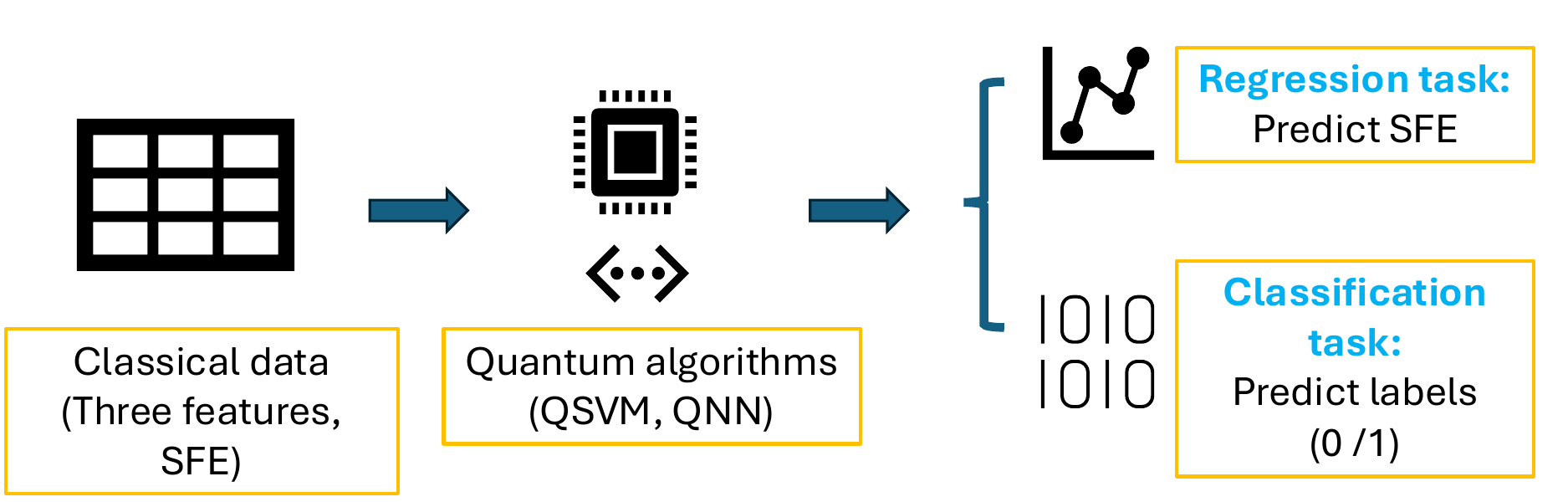}
    \caption{The workflow of this study. It consists of data preprocessing (e.g., normalization, preparation of training and validation data), data encoding into the quantum space and processing, model training/optimization, and validation.} 
    \label{fig:general}
\end{figure}

A general workflow of this study is shown in Figure \ref{fig:general}.
First, the classical data are preprocessed and tabulated in a structured form. The data consist of three features (bulk modulus $B$, elemental volume $V$, electronegativity $\nu$) and one label (stacking fault energies). In classification tasks, the label is converted into 0 and 1. When the value is larger than that of pure magnesium (19 mJ/m$^2$), then the label is 0; otherwise, the label is 1. The second step is to select one quantum algorithm, such as a quantum support vector machine or a quantum neural network. The classical data is first encoded into quantum space and then, depending on a specific algorithm, further processed to predict the labels. Similar processes are implemented for both regression and classification tasks.


\begin{figure*}
    \centering
    \includegraphics[width=1.0\linewidth]{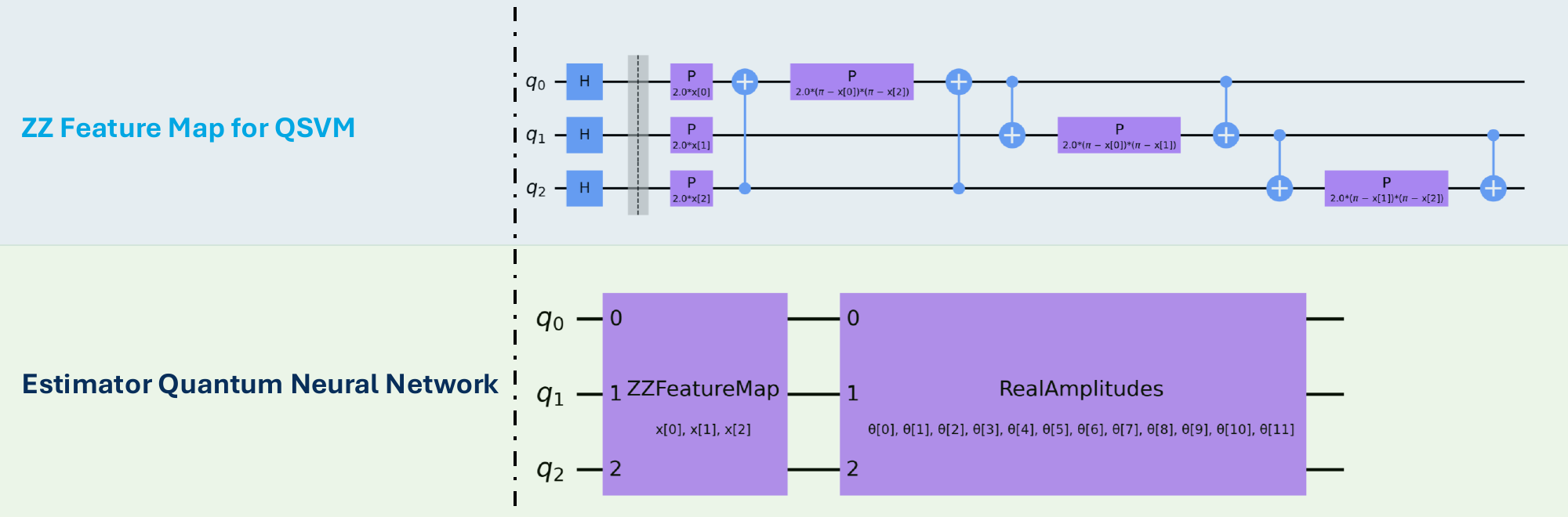}
    \caption{The two algorithms and their representative quantum circuits. Since the data in this study has three features, we only need three qubits. The first quantum circuit is the kernel in a quantum vector machine (QSVM). The second quantum circuit uses the quantum neural network (QNN) model.}
    \label{fig:algorithms}
\end{figure*}


To execute QML, we require the assistance of classical machine learning software and quantum software specifically designed for QML. Commonly used quantum computing software includes Qiskit \cite{javadiabhari2024quantumcomputingqiskit}, Cirq \cite{Cirq_Developers_2025}, Pennylane \cite{bergholm2022pennylaneautomaticdifferentiationhybrid}, TensorFlow Quantum \cite{broughton2020tensorflow}, etc. Our method contains both classical and quantum parts. Our quantum SVM and QNN models use Qiskit to manipulate quantum bits. The classical part relies on the SciKit-Learn \cite{scikit-learn_official} software.


\subsubsection{Quantum support vector machine}

The classical kernel of the SVM algorithm is replaced by a quantum kernel, which is computed by a quantum circuit.
We design a quantum circuit with multiple qubits and various quantum gates, such as the $n$-qubit quantum circuit in Figure \ref{fig:algorithms} (upper panel,~$n=3$). 
All the $n$ qubits are set to an initial state $|0\rangle^{\otimes n}$, and then various gates are added to manipulate the states. Standard quantum gates include rotation gates, Pauli gates, and, most importantly, the gates involving multiple qubits to introduce quantum entanglement. More specifically, we apply the unitary operation $U_{\Phi(\bm{x})}$ to the initial state for the data sample $\bm{x}$, where entanglement is included.
Then we have 
\begin{equation}
|\Phi(\bm{x})\rangle=U_{\Phi(\bm{x})}|0\rangle^{\otimes n},    
\end{equation}
where $U_{\Phi(\bm{x})} =\prod_{l}u_{\Phi(\bm{x})} H^{\otimes n}$. Here, $H$ is the Hadamard gate and
\begin{equation}
u_{\Phi(\bm{x})} =\exp\bigg{(} i\sum_{S\in [n]} \phi_S(\bm{x}) \prod_{i\in S} P_i \bigg{)}.    
\end{equation}
The quantum gates $P_i \in \{I, X, Y, Z\}$ are the Pauli gates \cite{havlivcek2019supervised}, and the index $S$ denotes the connectivity between qubits or data points. More details about the quantum gates can be found in Ref. \cite{pei2024designing}. Since $l$ determines the number of repetitions of the quantum gates $u_{\Phi(\bm{x})} H^{\otimes n}$, it is also referred to as the depth of the quantum circuit. The functions $\phi_{\{i\}}(\bm{x})=x_i$ and $\phi_{\{1,2\}}(\bm{x})=(\pi-x_1)(\pi-x_2)$ for $|S|=1$ and $|S|=2$, respectively. Here, the symbol $|S|$ represents the number of qubits involved.
We need $|S|=2$ to entangle the qubits. 
Assisted by quantum circuits, we map one data sample $\bm{x}$ in the training data to a quantum state $\Phi (\bm{x})$, i.e., $\bm{x} \in X \rightarrow |\Phi(\bm{x})\rangle \langle \Phi(\bm{x})|$. This procedure encodes the training data $X$ into the quantum space.
After the data encoding, quantum circuits can provide a quantum kernel $K(\bm{x},\bm{x}')=|\langle \Phi(\bm{x})|\Phi(\bm{x}')\rangle|^2$ needed for our quantum SVM models.

The quantum SVM algorithm adopted here is similar to the classical SVM algorithm except that quantum kernels or quantum circuits replace classical kernels in the classical SVM. So, we provide a brief introduction to the classical SVM and then describe how the quantum kernels and circuits are used.
The target of an SVM model is to find a hyperplane like $y=wX$ that meets the criterion
\begin{equation}
\mathrm{sign} (wX-y) ~\in [-1,1]
\end{equation}
for given dataset $\{X,y\}$,
where $w$ is a weight matrix.
The optimal hyperplane maximizes the separation of the data with different labels. The hyperplane perpendicular to the shortest distance between the two groups of subsets is an ideal choice. Mathematically, finding this hyperplane is equivalent to minimizing the reciprocal inverse magnitude of the weight vector $w$, i.e.,
\begin{equation}
    \min \bigg{\{} \frac{1}{||w||} \bigg{\}}.
\end{equation}

Before we perform the minimization, we need to choose a kernel $K(\bm{x},\bm{x}')$. The kernel maps the data from the original space to a new space. It measures the correlation between data points. In prediction, the kernel acts as a weight to determine the value $y$ for an unknown $X$. The closer the two points are, the higher the weight. The choice of a kernel is critical for the performance of an ML model. A well-chosen kernel can significantly ease the separation of different classes in this high-dimensional space. 
A kernel can be linear or non-linear, depending on the data distribution. For data with a complicated pattern, classical functions that can be expanded as polynomials may not be the best choice; instead, the quantum kernel, which incorporates quantum correlation and entanglement between qubits, offers a unique advantage. The guiding principle is that the more challenging it is to express the chosen kernel classically, the greater the benefit achieved. For example, a polynomial can hardly describe the quantum kernel with a high level of entanglement.
Figure \ref{fig:algorithms} (upper panel) shows an example of the quantum circuit used in our quantum SVM algorithm. We employ several different quantum circuits and tune their hyperparameters to find those that offer optimal performance. 
We have used the method in designing CCAs, where more qubits are involved \cite{pei2024designing}.

\subsubsection{Quantum neural network and hybrid model}

Different types of quantum neural networks (QNN) exist, such as Sampler QNN and Estimator QNN. QNN can be used in classification and regression tasks. In this study, we focus on the estimator QNN and optimize the hyperparameters involved. An example of Estimator QNN is shown in Figure \ref{fig:algorithms} (lower panel). Estimator QNN used a feature map like the entangled ZZFeatureMap \cite{javadiabhari2024quantumcomputingqiskit} to encode the classical data into the quantum space. This step is the same as QSVM. Differently, QNN has a higher level of involvement in quantum circuits. Unlike the hybrid QSVM, which outsources the rest of the algorithm to a classical algorithm, QNN continues to utilize another quantum ansatz (e.g., RealAmplitudes) to process the data and complete classification or regression tasks.

Like hybrid QSVM, there is a hybrid version of QNN where QNN becomes part of an ANN defined by PyTorch. In a hybrid QNN, the whole QNN circuit is used as a layer of the classical NN. More specifically, it wraps the QNN into a PyTorch module using TorchConnector. This can increase the nonlinearity of the NN and improve its accuracy. We will also use a hybrid QNN in this study.  
For more details of the hybrid model, please refer to \url{https://qiskit-community.github.io/qiskit-machine-learning/tutorials/05_torch_connector.html}. 







\subsection{Dataset}
The dataset used in this study is three elemental properties and the stacking fault energies (SFEs) when these elements appear in magnesium with a concentration of 6.25 at.\% \cite{sandlobes2014ductility}. The three elemental properties, including the bulk modulus ($B$ in gigapascals), atomic volume ($V$ in \AA$^3$), and electronegativity ($\nu$, unit-less), are taken as features, meaning a maximum of three qubits is needed. The SFEs are used as the target values in regression models. In classification models, we label all chemical elements with an SFE lower than 19 mJ/m$^2$ as 0 and the rest as 1.
The bulk modulus of Tc is missing in Ref. \cite{periodictable}, which is about 281 GPa from \url{https://www.matweb.com/search/datasheet_print.aspx?matguid=2629eceea53a4c0dbb9279fac6e6007d}



\begin{figure*} 
    \centering
    \includegraphics[width=1.0\linewidth]{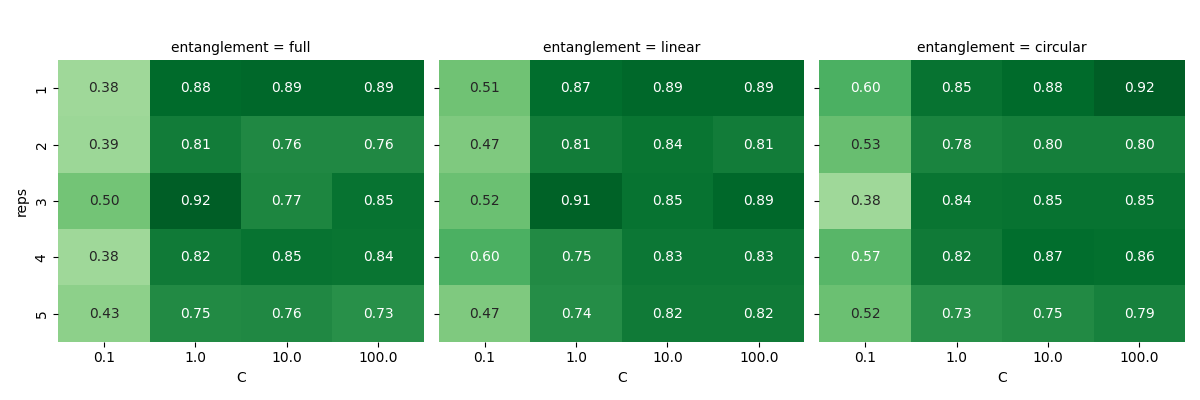}
    \caption{Validation scores of QSVC. Here, the validation scores with various $C$, $reps$, and $entanglement$ options are considered. }
    \label{fig:QSVC-heatmap}
\end{figure*}

\begin{figure*} 
    \centering
    \includegraphics[width=1.0\linewidth]{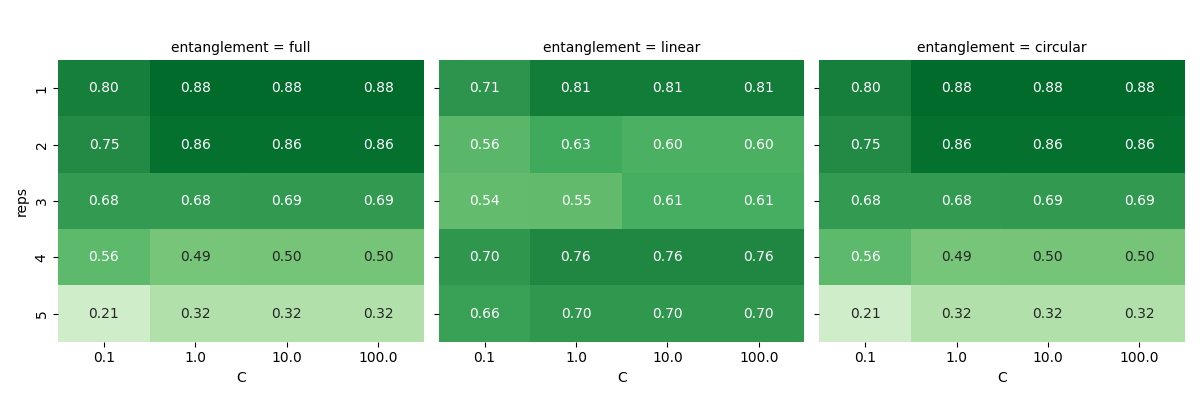}
    \caption{Validation scores of QSVR. Here, the validation scores with various $C$, $reps$, and $entanglement$ options are considered.}
    \label{fig:QSVR-heatmap}
\end{figure*}

\section{Results and discussion}


\begin{table} 
    \centering
    \caption{QSVC Hyperparameter Options}
    \label{tab:QSVC-parameters}
    \begin{tabular}{ccc}
        \toprule \toprule
        Hyperparameter & Type & Range\\
        \midrule
        C & Discrete & [0.1, 1, 10, 100] \\
        entanglement & Categorical & [circular, full, linear] \\
        reps & Discrete & [1, 2, 3, 4, 5]\\
        \bottomrule \bottomrule
    \end{tabular}
\end{table}

\subsection{Quantum SVM for classification}

We perform hyperparameter tuning and limit the range for a few important hyperparameters in Table \ref{tab:QSVC-parameters}. For this experiment, we consider 20 iterations for each model and adopt the common technique of five-fold cross-validations, i.e., shuffling train-test split, using 80\% of data for training and 20\% for validation. The final result is averaged over three tests. The hyperparameters included are listed below:
\begin{itemize}
    \item $C$: balances the accuracy of the model and the width of the margin in SVM. For example, a smaller value of $C$ means a larger margin and can tolerate more misclassifications. We examined a range of $C$ values corresponding to powers of 10, ranging from 0.01 to 100.

    \item reps: also called the depth of quantum circuit, usually represented by $l$. It is specific to quantum machine learning and means the number of repetitions for the quantum circuit. The values used are 1, 2, and 5. The value must be an integer and at least 1.

    \item entanglement: specific to the quantum machine library, and specifies which entanglement options to use. 
\end{itemize}

When $C$ is 1, 10, or 100, the accuracy is higher on average (\autoref{fig:QSVC-heatmap}). The value of gamma and the entanglement option do not show an obvious correlation with the test accuracy.
The optimal performance of 0.92 is achieved for $C=1$, reps=3, and full entanglement, which reduces marginally to 0.91 when we use the linear entanglement, keeping all other parameters identical. This indicates that a higher level of entanglement yields improved results, albeit weakly. A comparable performance is observed for $C=100$, reps=1, and circular entanglement.

\subsection{Quantum SVM for regression}

Similar to the classification task, we perform parameter tuning in C, reps, and entanglement to find out the most optimal parameter settings for the given task (Table \ref{tab:QSVR-parameters}). For this experiment, we compare the average result after 20 iterations of shuffled train-test split groups. A few parameters have been introduced above. There is one extra parameter in the regression task, epsilon or $\epsilon$:

\begin{itemize}
    \item $\epsilon$ or epsilon: defines a tolerance margin. For example, a smaller epsilon can be understood as using a smaller step size during model training, which generally yields more accurate results. We also checked $\epsilon$ for the exponential powers of 10, starting from 1 to 0.001.
\end{itemize}

\begin{table} 
    \centering
    \caption{QSVR Hyperparameter Options}
    \label{tab:QSVR-parameters}
    \begin{tabular}{ccc}
        \toprule \toprule
        Hyperparameter & Type & Range\\
        \midrule
        C & Discrete & [0.1, 1, 10, 100] \\
        epsilon & Discrete & [0.01, 0.001] \\
        entanglement & Categorical & [circular, full, linear] \\
        reps & Discrete & [1, 2, 3, 4, 5]\\
        \bottomrule \bottomrule
    \end{tabular}
\end{table}

From \autoref{fig:QSVR-heatmap}, we observe that performance in most cases is worse when $C=0.1$. In addition, the model performance drops at higher reps, accounting for circular and full entanglement, while the linear entanglement suggests a different trend.  
The optimal performance is found for reps=1 for the three types of entanglement, regardless of the choices for other parameters. More specifically, when we use circular or full entanglement, the optimal performance is 0.88 when $C=1, 10, 100$. The accuracy reduces to 0.8 when $C=0.1$. The accuracy decreases substantially to 0.81 when we switch the entanglement from full/circular to linear types. This apparent change clearly shows the importance of using high-level entanglement.

\subsection{Hybrid quantum neural network for classification}
\begin{figure*}
    \centering
    \includegraphics[width=0.8\linewidth]{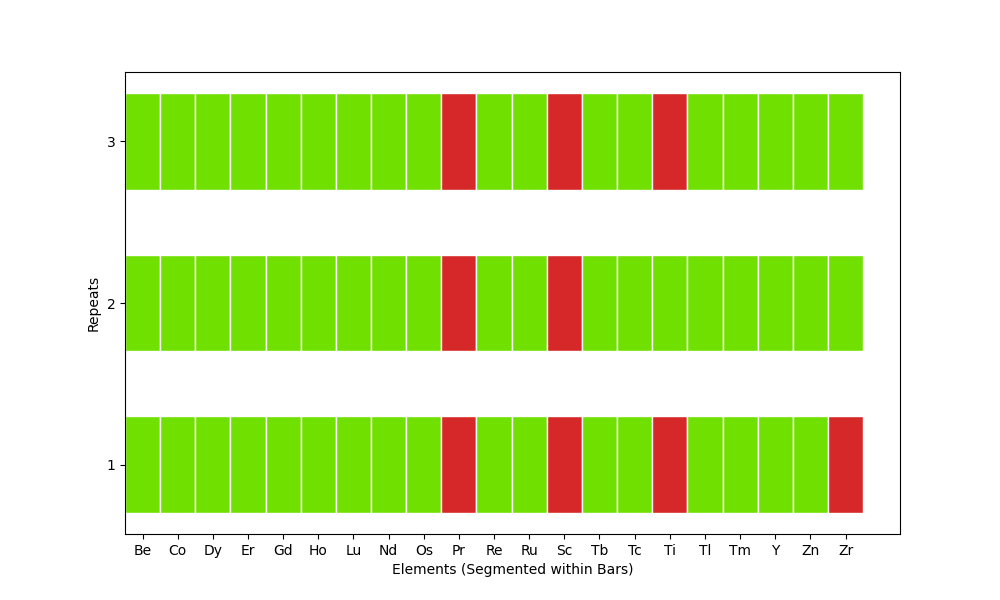}
    \caption{The hybrid QNN model for classification. The three horizontal bar graphs represent the correctness of the three repeats. Pr and Sc are wrongly predicted across all three repeats. The accuracy is 0.810 for $l=$ 1, 0.905 for $l=$ 2, and 0.857 for $l=$ 3.}
    \label{fig:QNN-classification}
\end{figure*}

We first use QNN on a classification task and consider key hyperparameters. A detailed implementation of the algorithm is provided in the Code Availability section.
Figure \ref{fig:QNN-classification} shows the QNN results for classification with various repeats or depths $l$ of core quantum circuits.
The three horizontal bar graphs represent the correctness of the three repeats. 
The accuracy is 0.810 for $l=$ 1, 0.905 for $l=$ 2, and 0.857 for $l=$ 3. These results indicate an optimal repeat of the core circuits $l=2$. Specifically, Pr and Sc are improperly predicted across all three repeats. When $l=1$, the labels of Ti and Zr are also improperly predicted. When $l=3$, Pr, Sc, and Ti are predicted with wrong labels. Therefore, the best performance or accuracy is obtained for $l=2$.

\subsection{Hybrid quantum neural network for regression}

\begin{figure}
    \centering
    \includegraphics[width=1.0\linewidth]{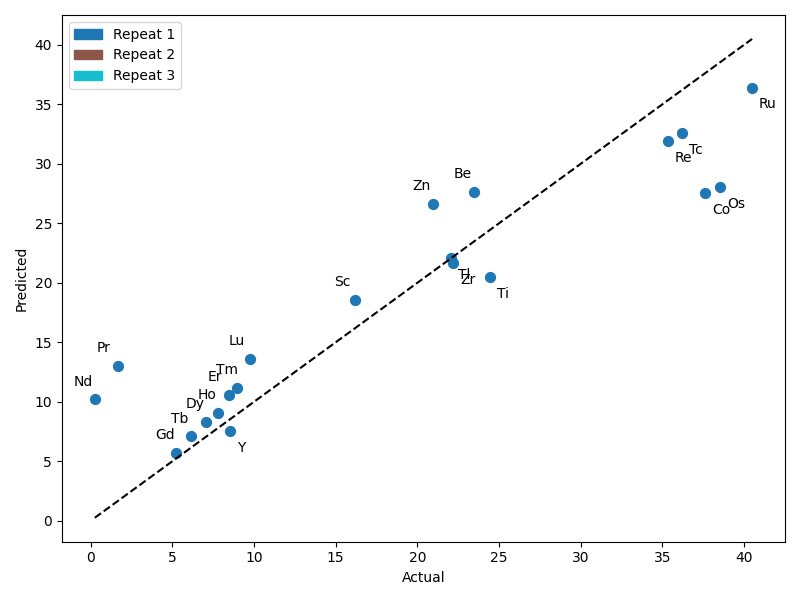}
    \caption{The hybrid QNN model for regression. Here, the horizontal and vertical axes represent the actual and QNN-predicted SFEs for different repeats or depths $l$ of the quantum circuit. Left side outliers: Nd, Pr, whose SFEs are the lowest; and right side outliers are Os and Co. Pearson's $R^2$ is 0.875 for repeat 1, 0.865 for repeat 2, and 0.864 for repeat 3.}
    \label{fig:QNN-regression}
\end{figure}

We adopted a hybrid model where the QNN model is wrapped in a PyTorch configuration for classical ML, allowing for better control over the neural network layers by adding an activation layer or customizing the loss function. We consider three different depths $l$ or a repeat of the quantum circuit: 1, 2, and 3.
The linear trend is evident in Figure \ref{fig:QNN-regression}. Overall, it follows the $y=x$ line with four chemical elements as outliers. Nd and Pr correspond to the smallest SFEs, making it difficult to predict; Os and Co are two non-rare earth elements that have SFEs larger than 35 mJ/m$^2$ and are predicted to be around 25 mJ/m$^2$. Since this number is larger than the SFE of pure Mg ($\sim$ 19 mJ/m$^2$), this prediction groups them into the solutes that cannot ductilize Mg, which is qualitatively correct.


\subsection{Future work}
We have implemented two types of QML algorithms in classification and regression tasks. Our goal for the future is to introduce more profuse quantum anzates or circuits in the context of machine learning, and systematically test their performance. These methods may or may not be applicable in QML, and their performance in solving real materials science tasks is unclear, making them valuable to the materials science community. 
One targeted algorithm is the variational quantum regressor/classifier, a representative method in the variational approach. The target is to minimize the error function $L(\bm{\theta})$ by finding the optimal parameters $\bm{\theta}$, where $L$ is defined by
\begin{equation}
\min L(\bm{\theta}), ~ L(\bm{\theta}) = \langle \psi|U^{\dagger}(\bm{\theta}) H U(\bm{\theta})|\psi \rangle,
\end{equation}
where the parameters $\bm{\theta}$ will be determined at minimal error function for all given wavefunctions.
One part of the quantum circuit is used to implement the parameters $\bm{\theta}$, while a separate part encodes the input data $X$.
To accelerate quantum computing, either in quantum machine learning or quantum chemistry simulations, GPU acceleration is implemented, where multiple quantum circuits corresponding to different $\bm{\theta}$ values are computed simultaneously.

The ultimate goal of this series of studies is to identify the common features or patterns that are likely to lead to successful high-performance QML models for practical tasks rather than idealized, simplified tasks based on synthesized data.

\section{Conclusions}

Quantum machine learning models are usually tested on synthetic or idealized data in previous studies.
Using data from a real materials-science problem, we examine the performance of two quantum algorithms: one quantum algorithm and one hybrid quantum algorithm, specifically the quantum support vector machine and the Estimator quantum neural network. We developed models for both classification and regression tasks using these algorithms.
In the quantum support vector machine for the classification task, an optimal accuracy of 0.92 is achieved by three fully entangled qubits in a quantum circuit with a depth of $l=3$. In the quantum support vector machine for the regression task, an optimal accuracy of 0.88 is achieved by three fully or circularly entangled qubits in a quantum circuit with a depth of $l=1$.
In the hybrid quantum neural network model for the classification task, an optimal accuracy of 0.91 is achieved by a three-qubit quantum circuit with a depth of $l=1$. In the quantum support vector machine for the regression task, an optimal accuracy of 0.88 is achieved by a three-qubit quantum circuit with a depth of $l=1$. 
Our results confirmed that even with the same data and algorithm, the best model is obtained with very different hyperparameters. This result confirms the complicated fact that we need to tailor the hyperparameters to get an optimal model, even for the same data and algorithm in different tasks. Overall, complex entanglement states, either circular or fully entangled, are superior to non-entangled states.
The promising accuracies of these four models indicate their applicability in real materials science problems.
Our study offers valuable insights into the applications of quantum machine learning in addressing real-world problems.

\section*{Code availability}
The code used in this study is available on GitHub:
The version of code that uses five-fold cross-validations: \url{https://github.com/lw3266/QML-SFEPrediction}; the version of code that uses twenty-fold cross-validations: \url{https://github.com/gongyilun/QMML_SFE_Mg/tree/main}.
To help reproduce the results, we provide a list of the specific packages and their versions below:
first create a python environment with \textit{Python 3.11.4};
then, pip install \textit{qiskit==0.46.3 qiskit-aer==0.12.1 qiskit-algorithms==0.3.1 qiskit-ibm-provider==0.6.1 qiskit-machine-learning==0.7.2 qiskit-terra==0.46.3
qiskit                    0.46.3
qiskit-aer                0.12.1
qiskit-algorithms         0.3.1
qiskit-ibm-provider       0.6.1
qiskit-ibmq-provider      0.20.2
qiskit-machine-learning   0.7.2
qiskit-terra              0.46.3}.













\end{document}